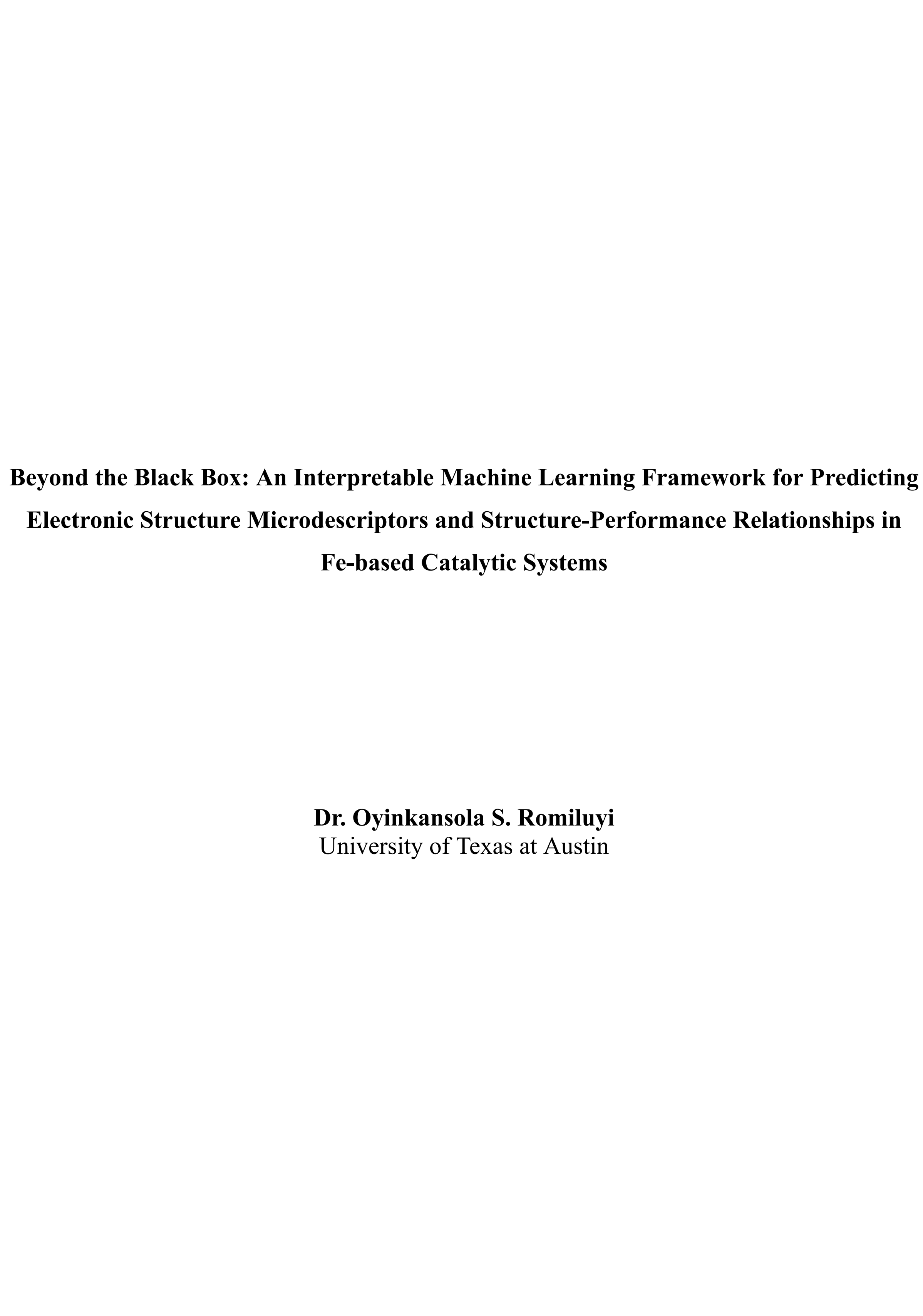

# Beyond the Black Box: An Interpretable Machine Learning Framework for Predicting Electronic Structure Microdescriptors and Structure-Performance Relationships in Fe-based Catalytic Systems

**Dr. Oyinkansola S. Romiluyi**
University of Texas at Austin

**Abstract**

The current catalyst discovery, design, and development pipeline for applied science and engineering applications is still largely conducted via brute force, involving extensive and expensive trial-and-error experimentation by scientists with decades-long yet irreproducible chemical intuition and expertise. This pipeline remains a major, central bottleneck in the deployment of scalable solutions in various energy- and capital-intensive applications like efficient methane conversion technologies. Despite major advances in *ab initio* simulations like DFT, the publication of open catalysis datasets and literature studies, and improved high-throughput computation geared towards catalyst discovery, a generalized, data-driven framework that explicitly links catalyst performance to the notoriously complex catalyst design space is not yet widely available. Moreover, the various catalyst properties documented in literature for a multitude of reactions have hardly been quantified in terms of their relative importance for said reaction microenvironments. By using publicly-available catalyst datasets and repositories (e.g., The Materials Project), this work presents a machine learning workflow that combines SHAP-based feature importance analysis (i.e., explainable AI or xAI) with non-linear, tree-based models (Random Forest and Bayesian-optimized CatBoost) that are trained to assess the black-box system of choice: Fe-zeolite and oxide-supported catalysts (e.g., Fe ZSM-5, Fe/Si/Al/O) used in the catalytic partial oxidation of methane (i.e., POM). With a small dataset and limited evaluations, those microdescriptors or features (i.e., thermodynamic, structural, geometric) that strongly influence the electronic band gap and that most likely govern macroscale, end-use catalytic performance (i.e., catalyst selectivity, activity, stability, turnover frequency) are uncovered and ranked in order of impact on the predicted target variable. Thus, this study successfully yields a generalizable machine learning framework and a prioritized set of physically meaningful features that crucial, early-stage R&D work should focus on for accelerated catalyst screening and discovery. When subsequently combined with microkinetic models, these quantifiable features can be converted into selectivity and conversion metrics and input directly into reaction engineering models that simulate real-world POM reactor systems, effectively creating an end-to-end digital twin of a complex technology that can be used for optimization before any bench or pilot test need be conducted.

## Table of Contents

## 1. Introduction

### *1.1 Background and Broader Context*

Catalysis is a core foundation of our modern society, with over 90% of chemical manufacturing relying on various catalytic processes that impact at least 35% of the global GDP across various industries: fuels, fertilizers, pharmaceuticals, polymers, and environmental mitigation technologies [1]. The widespread use of catalysts enables chemical transformations that would otherwise be too slow, energy-intensive, or environmentally costly to implement at scale, such as ammonia synthesis (i.e., the Haber-Bosch process), petroleum refining, and olefin production (e.g., the Fischer-Tropsch (FT) process). It also determines the feasibility and efficacy of environmental remediation processes, emissions control technologies, and emerging pathways for carbon capture and utilization, as well as for hydrogen, green ammonia, and sustainable fuels production. Asides from promoting these key thermal conversion reactions, catalysis also plays a critical role in industrial electrochemical or electrocatalytic processes and advanced materials manufacturing, such as in the chlor-alkali process for chlorine production, in the water electrolysis process used in fuel cells, in battery manufacturing, and in various electroplating methods used in metals refining operations [2]. As the global energy, chemicals, and materials landscape gradually transitions toward more low-carbon, circular, and energy-efficient systems, the importance of catalysis is only expected to increase [3]. Thus, understanding catalysis has vast, global technological, economic, and societal ramifications in the realms of energy security, environmental preservation, and industrial commercial competitiveness.

### *1.2 The Foundations and Complexity of Modern Catalysis*

Fundamentally, catalysis is about lowering the energetic barriers that govern chemical transformations or reactions. In heterogeneous systems, where the catalyst is in a different physical state than the reactant environment it is operating in (e.g., solid catalyst pellets used in gaseous reactions), this energy barrier manipulation usually depends on how well the catalyst surface facilitates the chemical bond activation of the reactant or adsorbate molecules and the subsequent bond-formation and/or bond-breaking steps. For example, methane ($CH_4$) activation is notoriously difficult because the first C-H bond requires ~413 kJ/mol of energy to cleave or break and the resulting selectivity of valuable oxygenate products like methanol, ethanol, and other alcohols can be hard to maximize. In fact, only certain catalyst active site electronic environments, typically those involving iron (Fe), nickel (Ni), cobalt (Co), copper (Cu), or platinum (Pt) transition metal or d-block metal families in the periodic table, enable this methane chemical transformation at industrially-relevant temperatures and pressures [4]. Catalytic reactions go through cycles of reactant molecule adsorption, surface reaction, and desorption. For various chemistries, the core challenge is in achieving the right balance between an adsorption that is strong enough to activate chemical bonds in the reactants but not so strong such that the reaction intermediates overbind to and accumulate on the catalyst surface, which can cause undesirable effects like coking or overoxidation in the presence of oxygen. This delicate and complex balance of binding energy is classically called the "Sabatier principle" which states that the optimal catalyst binds intermediates neither too weakly nor too strongly, giving rise to the widely observed "volcano" relationships in catalysis (i.e., catalytic activity or reaction rate is plotted against a microdescriptor of the binding strength (e.g., heat of adsorption), with the catalyst with the highest activity being at the peak of the volcano plot) [5], [6].

Underlying many of these aforementioned concepts is the "d-band theory" in catalysis, which correlates the electronic structure of transition or d-block metals (i.e., the d-band center) to their catalytic activity or behavior. Specifically, d-band theory describes the delicate adsorption-desorption balance and strength of the chemical bond between the transition metal catalyst surface and a reactant/adsorbate molecule via the metal's d-band center (i.e., a single energy level that represents the position and electron occupancy/number of electrons filling the metal's d-orbitals or the average energy of the metal's d-orbitals) relative to the Fermi level (i.e., the energy level of the highest occupied electrons in the catalyst or the thermodynamic work required to add one electron to the system) [7]. A d-band center that is close to the Fermi level generally indicates a stronger interaction with reactants, higher adsorption energy, and therefore higher catalytic activity. But, as discussed before, an optimal moderate position that is neither too strong nor too weak is often more ideal. Thus, the d-band model provides a useful framework for predicting and understanding trends in the catalytic activity of transition metal catalysts and manipulating the d-band center is therefore crucial for the rational design of better catalysts for specific reactions. For instance, doping the transition metal with other elements can change the electronic structure and shift the d-band center while implementing various material science property techniques like strain engineering, alloying, and promoting the presence of defects can also shift the d-band center; these will affect the adsorption and activation of specific reactant intermediates and thus alter catalytic performance [8]. This is where the electronic structure microdescriptors of catalysts become useful, a few of which are:

- band gap: the energy difference between a material's valence and conduction bands
- formation energy: a material's or reaction intermediate's stability on a catalyst surface, used in first-principle or *ab initio* computational studies (e.g., density functional theory (DFT))
- coordination number: a structural descriptor for a metal atom that dictates its reactivity
- electron density: the distribution of electrons in a molecule or on a catalyst surface
- energy above hull: a thermodynamic measure of a material's stability

These microdescriptors represent and summarize the complex electronic structure relationships involved in catalysis into singular numbers that can be correlated to a reaction's activation energy, reactant mobility, radical stability, redox (i.e., reduction and oxidation) flexibility, and lattice stability, and subsequently connected to macroscale figures-of-merit that assess catalytic performance like conversion, selectivity, yield, and turnover frequency (TOF) via microkinetic rate laws [9], [10]. Most of the chemical intuition surrounding this complexity is currently gained through years of trained research and literature study paired with extensive experimentation and empirical observation, notwithstanding the fact that each reaction system, microenvironment, and catalyst of interest has its own inherent complexities that require further dedicated study.

### *1.3 The Catalytic Methane Partial Oxidation Reaction (POM)*

Methane ($CH_4$) is the primary component of natural gas and biogas and large quantities are used yet wasted through flaring, venting, and leakage in countless industrial daily operations. As mentioned previously, direct conversion of methane into value-added products of interest like methanol or syngas is especially difficult because it is an inherently stable molecule, its C-H bond strength makes activation kinetically demanding, and the products of interest are often thermodynamically more prone to further, undesirable overoxidation to carbon dioxide ($CO_2$) than methane itself [11]. Among proposed conversion routes, the catalytic partial oxidation of methane (i.e., POM) has attracted sustained interest because it offers a direct pathway to value-added products, most notably syngas (i.e., a mixture of carbon monoxide (CO) and hydrogen ($H_2$)), with more favorable thermodynamics, compact reactor design, and lower energy input compared to steam methane reforming (i.e., SMR) [12].

$$CH_4 + 1/2O_2 \rightarrow CO + 2H_2$$

*Equation 1. Partial oxidation of methane (POM) to syngas.*

Over multiple decades, a wide range of catalyst families have been explored for POM, including noble metals like rhodium (Rh), platinum (Pt), and palladium (Pd), perovskites, and various transition metal oxides [13], [14]. Rh-based catalysts and similar noble metals have historically shown the highest activity and stability, but are usually undesirable due to their high cost and supply limitations [15]. On the other hand, iron/Fe-based systems, particularly Fe-zeolites (i.e., Fe/Si/Al/O), which are aluminosilicate compounds where Fe has been doped into a microporous zeolite structure via ion exchange, wet impregnation, or direct synthesis/incorporation, represent a promising alternative. They can selectively activate methane and oxygen, stabilize reactive species and intermediates (e.g., $CH_3$*, O*, OH*, CO*, $CO_2$*) , are relatively low cost, have high thermal stability, and benefit from tunable, isolated active sites and shape-selective microenvironments that impose geometric constraints, control diffusion/transport phenomena, and confine reactive intermediates to allow for the suppression of overoxidation to $CO_2$ [16].

But, regardless of catalyst choice, POM involves complex surface chemistry in the cleavage and controlled activation of strong C-H bonds under oxidizing conditions, a regime in which catalysts must balance two competing outcomes: enabling rapid activation of methane while suppressing overoxidation to $CO_2$. Its reaction microenvironment and macroscale performance therefore place stringent demands on and is very sensitive to (1) the adsorption energy balance which influences the surface coverage and catalyst activation barriers, (2) the d-band center which correlates with bond strength and scaling relations that dictate the binding energies of different reaction intermediates on a catalyst surface, and (3) the various aforementioned electronic structure microdescriptors. This all makes POM both a chemically-rich and industrially-relevant system for studying electronic structure-performance function relationships and for pursuing accelerated performance predictions, optimizations, and catalyst discovery initiatives as an active field of study.

### *1.4 Using Machine Learning for Scientific and Catalyst Discovery*

The catalyst discovery research and development (R&D) pipeline remains a central constraint in chemical process design because most industrial transformations operate in complex and challenging reaction environments where underlying structure-function relationships are only partially and explicitly understood, with the aforementioned catalytic partial oxidation of methane (i.e., POM) being a prime, representative example. Traditional development workflows tend to rely heavily on domain expertise, experience, and intuition, slow trial-and-error iteration and experimentation, and low-throughput screening. The catalyst design space can also easily grow combinatorially large with choices of metals, supports, dopants, particle structures, facets, surface terminations, and operating conditions, causing the number of plausible catalyst candidates to quickly exceed what any laboratory can reasonably test experimentally, even over many years of experimental trials. Recent advances in materials informatics have demonstrated that machine learning (ML) models can extract meaningful structure-property relationships from large datasets of DFT calculations and experimental measurements, with several studies using neural networks or gradient-boosted trees to predict adsorption energies or reaction energetics on metal and oxide surfaces [17], [18], [19]. And various publicly-available catalyst datasets like the Materials Project [20], the Open Catalyst Project [21], Catalysis-Hub [22], NIST Chemical Kinetics Database [23], and related repositories have enabled supervised learning of bulk properties and adsorption energies across thousands of surfaces and catalyst compositions. These dual efforts have shown that non-linear models can be used to predict adsorption energetics at a fraction of the time and computational cost of traditional DFT and can support the rapid screening of reaction pathways and catalytic surfaces. Parallel work has also applied ML directly to catalyst discovery: Ulissi et al. and Suvarna et al. introduced active-learning schemes for electrocatalyst screening [24], [25] while Gómez-Bombarelli et al. integrated generative models with property predictors for molecular and materials design [26]. More recent work has extended these approaches to identify promising high-entropy alloys [27], optimize catalyst compositions with Gaussian-process surrogate models [28], optimize processing conditions for metal halide perovskites [29], and support chemical discovery in autonomous laboratories [30], [31], [32].

In particular, Bayesian optimization (BO) has emerged as a key tool in these catalyst design and discovery workflows since it is a sample-efficient technique widely used for optimizing expensive objective functions where evaluations are limited by time, cost, or computation, which makes it well-suited for R&D settings where only a small number of synthesis or testing iterations are feasible [33], [34]. In chemistry, materials science, and catalysis, BO has been applied to reaction condition optimization, formulation search, electrocatalyst screening, and closed-loop materials discovery, emphasizing its efficiency in navigating high-dimensional, noisy design spaces with very few evaluations (typically using Gaussian-process or tree-based surrogate models) which often underpins autonomous/semi-autonomous experimentation platforms that couple robotics and in-line analytics [35], [36], [37], [38], [39], [40]. A second emerging area is interpretable ML or explainable AI (i.e., xAI): while many catalyst ML studies report accurate predictions, fewer provide transparent, microdescriptor-level explanations that can be reconciled with physical, chemical, and/or mechanistic understanding. SHAP interpretability (i.e., SHapley Additive exPlanations), developed by Lundberg and Lee, has recently been adopted in materials science because it produces model-agnostic, directionally meaningful feature contributions or attributions that align with existing chemical intuition [41]. Recent work has used SHAP to interpret adsorption energy predictions, stability trends, and surface reactivity in alloy and oxide materials systems and in photocatalysis [42], [43].

### *1.5 Open Research Questions and Motivation For Current Study*

Despite the significant recent progress described in the above literature review, a few notable gaps remain. First, most prior studies use large task-specific datasets (for example, adsorption energies for a single reaction mechanism) rather than evaluating how much insight can be extracted from more general-purpose materials microdescriptors, such as formation energy, band gap, elemental ratios, density, and site counts. These microdescriptors are widely available, but are often underutilized in early-stage catalyst screening. Second, while many studies have trained predictive models for catalyst discovery, far fewer have attempted to pair model interpretation with Bayesian optimization in a unified workflow that ranks which microdescriptors matter most and fundamentally dictate catalytic behavior for a given reaction system. Such integration is increasingly relevant for autonomous laboratories and digital-twin frameworks, where mechanistic interpretability and sample efficiency are both required for intelligently exploring the complex catalyst design space so that the allocation of limited lab resources can be determined *a priori*.

This study addresses these gaps via a combined ML approach: using supervised learning to train, benchmark, and quantify microdescriptor interactions in linear (Linear Regression) and non-linear tree-based (Random Forest and Bayesian-optimized CatBoost) models on a publicly-available Fe-zeolite (i.e., Fe/Si/Al/O) dataset combined with SHAP interpretability and feature importance analysis to rank which microdescriptors strongly govern band gap variation and likely exert the strongest influence on catalytic outcomes. Band gap is chosen as the target variable because it is a known proxy for electron transfer, redox behavior, and methane and oxygen reactant bond activation energies in oxide-based catalysts for POM [44]. It is also correlated with the d-band center since both are determined by atomic orbital overlap and shifting the d-band center can lead to changes in band gap and overall electronic structure. But these two variables are likely non-linearly correlated since other factors, such as the presence of defects or the nature of the complex catalytic microenvironment, can cause interference [45], [46], [47]. The contribution of this work is thus twofold: (1) it shows that a modest set of structural and thermodynamic microdescriptors commonly available in large materials databases can still uncover physically meaningful electronic structure-property relationships when paired with interpretable ML or explainable AI techniques and (2) it demonstrates that a lightweight, integrated SHAP and BO framework can provide an efficient and scientifically grounded foundation for prioritizing catalyst microdescriptors towards autonomous catalysis R&D.

## 2. Datasets and Methodology

This study was designed to be a compact R&D screening pipeline using supervised ML, SHAP and feature importance analyses, and Bayesian optimization to analyze and quantify microdescriptor-performance or (electronic) structure-performance relationships relevant to the Fe-based catalyst, methane partial oxidation (i..e, POM) reaction system. It starts from a real-world, structured dataset of catalyst microdescriptors, builds a family of three supervised learning linear and non-linear tree-based models, and then interprets the learned structure-performance relationships via SHAP feature importance analysis. All analyses were conducted in Python 3 in a Google Colab cloud notebook environment designed to be version-controlled via Git and using the pandas and NumPy, scikit-learn, CatBoost, SHAP, scikit-optimize/skopt, and matplotlib libraries for data manipulation, baseline modeling and metrics, gradient-boosted trees, feature attribution, hyperparameter tuning, and generating plots, respectively.

### *2.1 Data Sources, Data Preparation, and Exploratory Analysis*

Options of open datasets included the Open Catalyst Project (OCP), the Materials Project, Catalysis-Hub, and the NIST Chemical Kinetics Database from NIST Library. However, the Materials Project was specifically chosen out of all of these eligible datasets for the versatility of accessing and pulling online data programmatically using an open license Materials Project API key, rather than downloading dated CSV files. Querying the database for relevant ternary and binary Fe-containing compounds (i.e., Fe/O/Al, Fe/Si/Al, Fe/O etc.) returned a raw dataset containing just under 300 samples with 13 associated microdescriptors which would constitute the feature list: formation energies, per atom metrics, energy above hull, elemental fractions and stoichiometric ratios, site counts, and geometric measures like volume and density, with band gap (in eV) chosen as the target property variable. Data preparation and cleaning followed a minimal-processing approach where instances with incomplete or missing target values and obvious data errors and duplicates were removed and where continuous variables/features were preserved in their original, unscaled physical units in order to maintain physical interpretability. No categorical variables/microdescriptors were used in this present study. The preprocessed dataset was then randomly partitioned into a standard 80/20 train-test split, with the test set strictly reserved for final model evaluation: no information from the test set was used during feature scaling, (3-fold) cross-validation, or hyperparameter tuning. Principal component analysis (PCA) was used only as an exploratory, diagnostic tool to examine the microdescriptor covariance structure/distribution and to provide a compact summary of correlated variables. Feature engineering was deliberately modest and limited to simple, physically interpretable transformations rather than automated feature expansions or polynomial transformations. All raw energetic microdescriptors were retained because they were expected to reflect interactions relevant to bond formation, bond strength, and general electronic structure.

### *2.2 Modeling Framework, Bayesian Optimization, and SHAP Interpretation*

This study treats methane partial oxidation (POM) catalytic performance as a regression problem with up to 13 features/independent variables and with band gap as the chosen target variable. Ordinary least-squares Linear Regression served as the low-complexity baseline model while Random Forest and the Bayesian-optimized, gradient-boosted decision tree CatBoost regressor served as medium- to high-complexity non-linear models of interest for the resulting small-sized dataset. All models were evaluated on the test set using the following standard metrics to assess and compare regression performance: mean absolute error (MAE) as a measure of prediction error in physical units, root mean squared error (RMSE) which emphasizes larger errors, and the coefficient of determination ($R^2$) to indicate how much of the variance in the target variable a chosen model explains. Parity and residual plots were also generated to diagnose any systematic prediction errors and assess whether any model exhibited notable bias, overfitting, or underfitting across the band gap range. Interpretability analysis was then conducted using feature importance and global SHAP ranking to quantify the contribution of each microdescriptor to the non-linear tree-based model predictions over their respective domains.

## 3. Results and Discussion

### *3.1 The Band Gap Landscape Of Fe/Si/Al/O Materials*

Across the just under 300 Fe/Si/Al/O entries extracted from the dataset, band gaps spanned from 0 eV (indicative of metallic or near-metallic materials behavior) up to 5.749 eV (more indicative of insulating and semiconducting materials). This empirical distribution is highly right-skewed, with most of the materials (>50%) possessing near-zero band gaps and clustered below 0.0438 eV and with a long, sparse tail extending to higher energies (Figure 1).

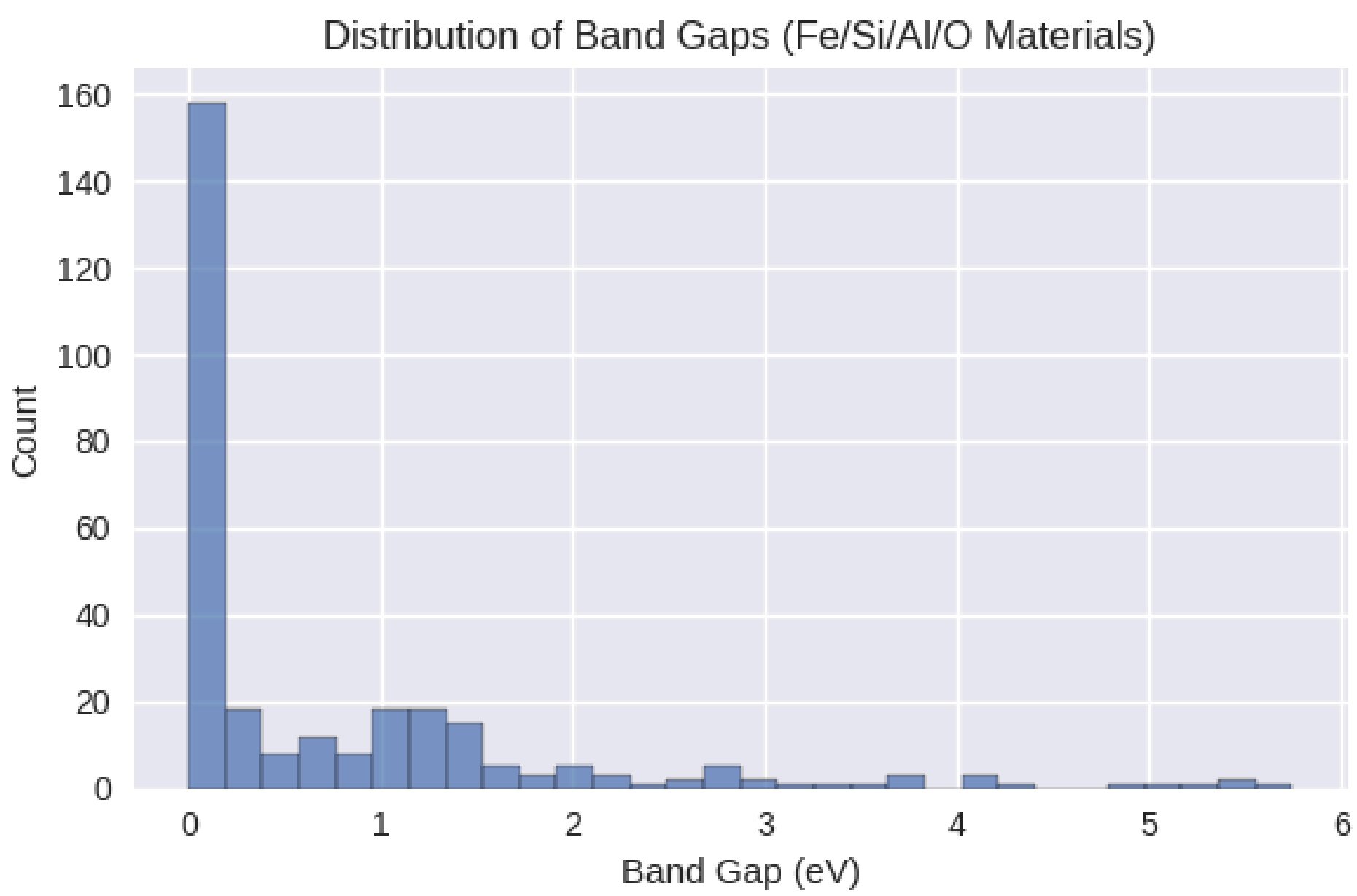

*Figure 1. Band gap distribution for the 297 Fe/Si/Al/O materials considered.*

This target variable distribution pattern indicates that there is substantial electronic discrimination within the mixed-valence Fe/Si/Al/O design space. Interestingly, a more narrow or low band gap characteristic is known to support easier electron transfer or electron hopping (via intervalence charge transfer (IVCT)), which is the mechanism responsible for the enhanced electrical conductivity observed in metals [48]. From a catalysis standpoint, this is desirable since this favors high redox flexibility and strong activation of $O_2$ and $CH_4$, but this can risk overoxidation. Wide or higher band gap frameworks are more insulating, have lower redox ability, and tend to suppress charge transfer; this can potentially lead to more controlled radical chemistry that can protect against overoxidation during partial oxidation but it can also reduce overall catalytic activity [49]. This band gap tradeoff means that the ensuing learning task is nontrivial and that band gap is indeed a meaningful and suitable surrogate metric for studying Fe-based methane partial oxidation catalysts.

### *3.2 Microdescriptor Structure*

Principal component analysis (PCA) on the 13 structural and energetic microdescriptors shows that the first three components capture and explain ~63% of the total variance (Figure 2). This indicates that no singular microdescriptor dominates variance: PC1 accounts for 30.2% of variance and is strongly associated with the formation energy per atom and energy per atom microdescriptors, PC2 (18.7%) captures geometric parameters such as volume, density, and site count, while PC3 (14.4%) incorporates compositional/stoichiometric fractions of Fe, Si, Al etc.

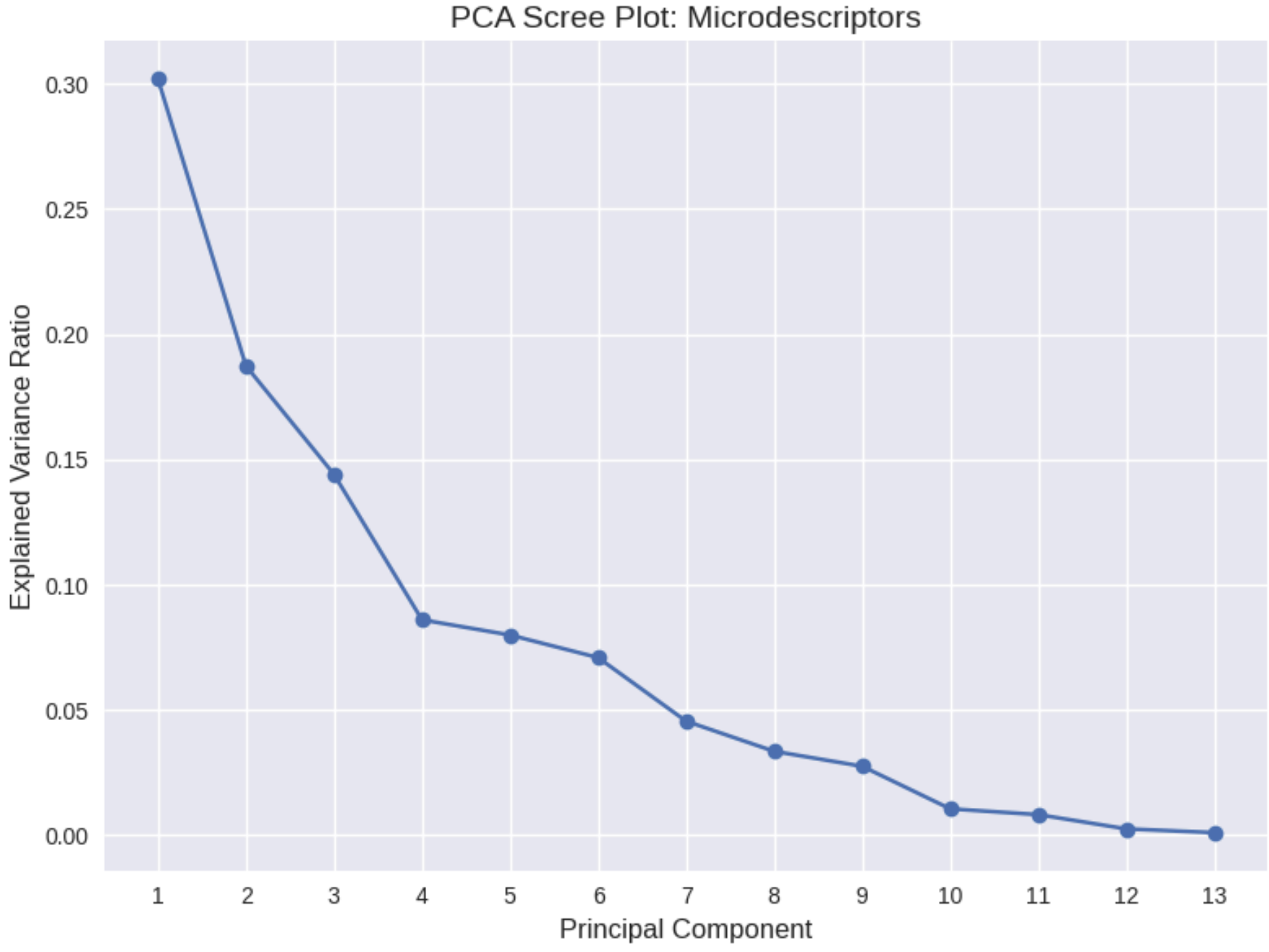

*Figure 2. PCA scree plot showing the variance ratios for the (first) thirteen principal components.*

Plotting the PCA scores with band gap as a color map in a PC1-PC2 scatterplot showed a moderate separation in how different electronic regimes occupied the microdescriptor space (Figure 3). Materials with larger band gaps tended to fall in regions associated with more open, expanded geometric frameworks, more metallic structures clustered mostly oppositely in the map, while materials with intermediate band gaps laid in between these two groups.

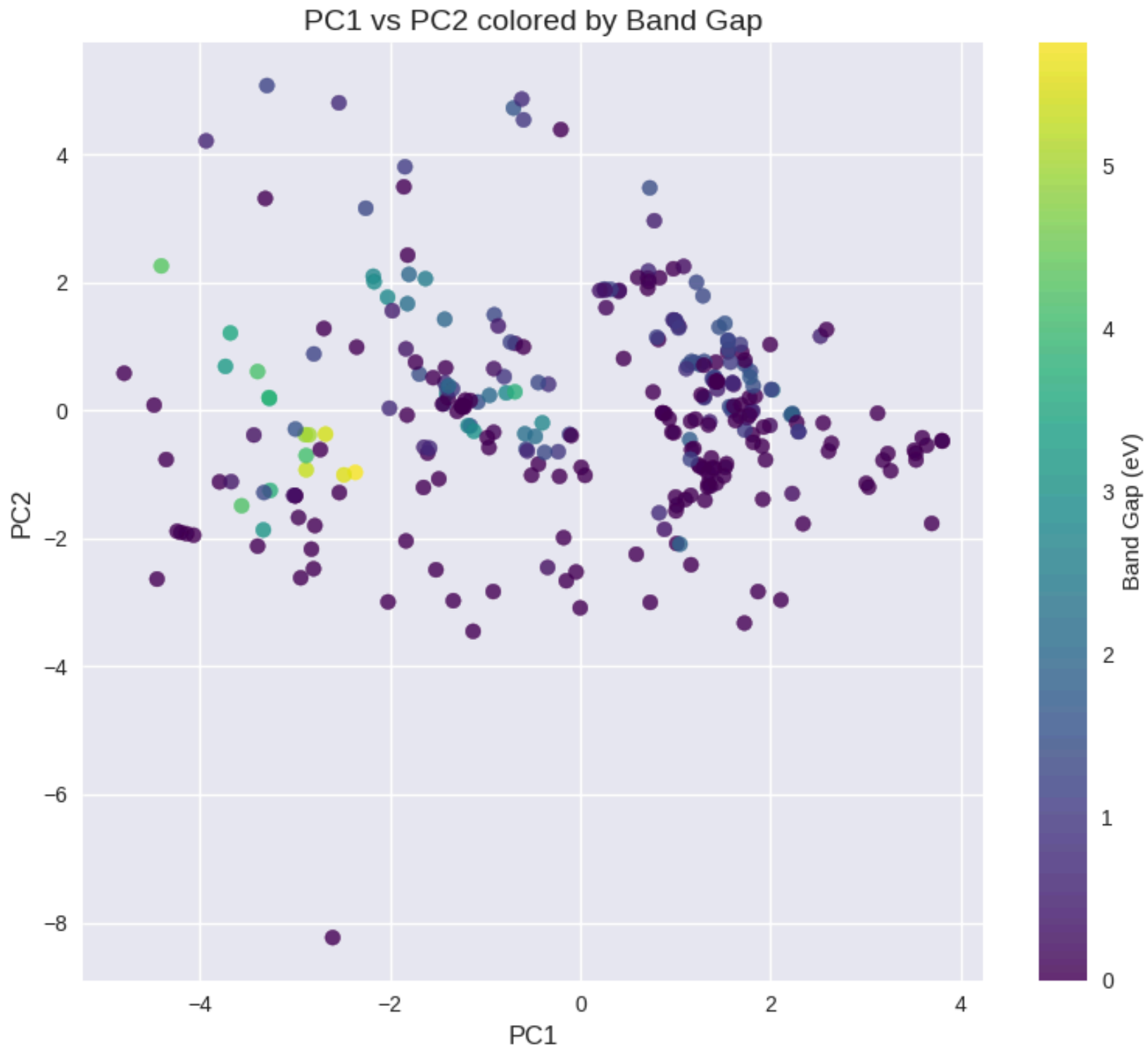

*Figure 3. Projection of the dataset onto the first two principal components, colored by band gap.*

This pattern indicates that relatively simple combinations of thermodynamic and geometric microdescriptors already separate major classes of electronic behavior [50], [51]. In other words, the Fe/Si/Al/O landscape has a low-dimensional structure that can be reasonably exploited by supervised models without having to construct highly engineered features.

### *3.3 Model Performance Comparisons*

Comparing the three models, Linear Regression, Random Forest, and Bayesian-optimized CatBoost, shows that introducing non-linear, tree-based models improved predictive performance substantially (Figure 4). Linear Regression predictably performed poorly with an MAE of ~0.75 eV, RSME of ~0.97, and an $R^2$ of ~0.32, which is consistent with the fact that band gap is not expected to be a linear function of formation energy, geometric properties, or elemental fractions and that the microdescriptors are expected to interact in non-linear ways. The Random Forest and CatBoost models both appear to capture much more of these non-linearities and demonstrate stronger performance: Random Forest achieves an MAE of ~0.39 eV, RSME of ~0.56, and an $R^2$ of ~0.77 while the Bayesian-optimised CatBoost model reached an MAE of ~0.44 eV, RSME of ~0.61, and an $R^2$ of ~0.73. In this particular dataset, Random Forest slightly outperformed CatBoost, likely because the feature space is low-dimensional and relatively clean, so the extra flexibility of gradient boosting did not offer a clear generalization advantage.

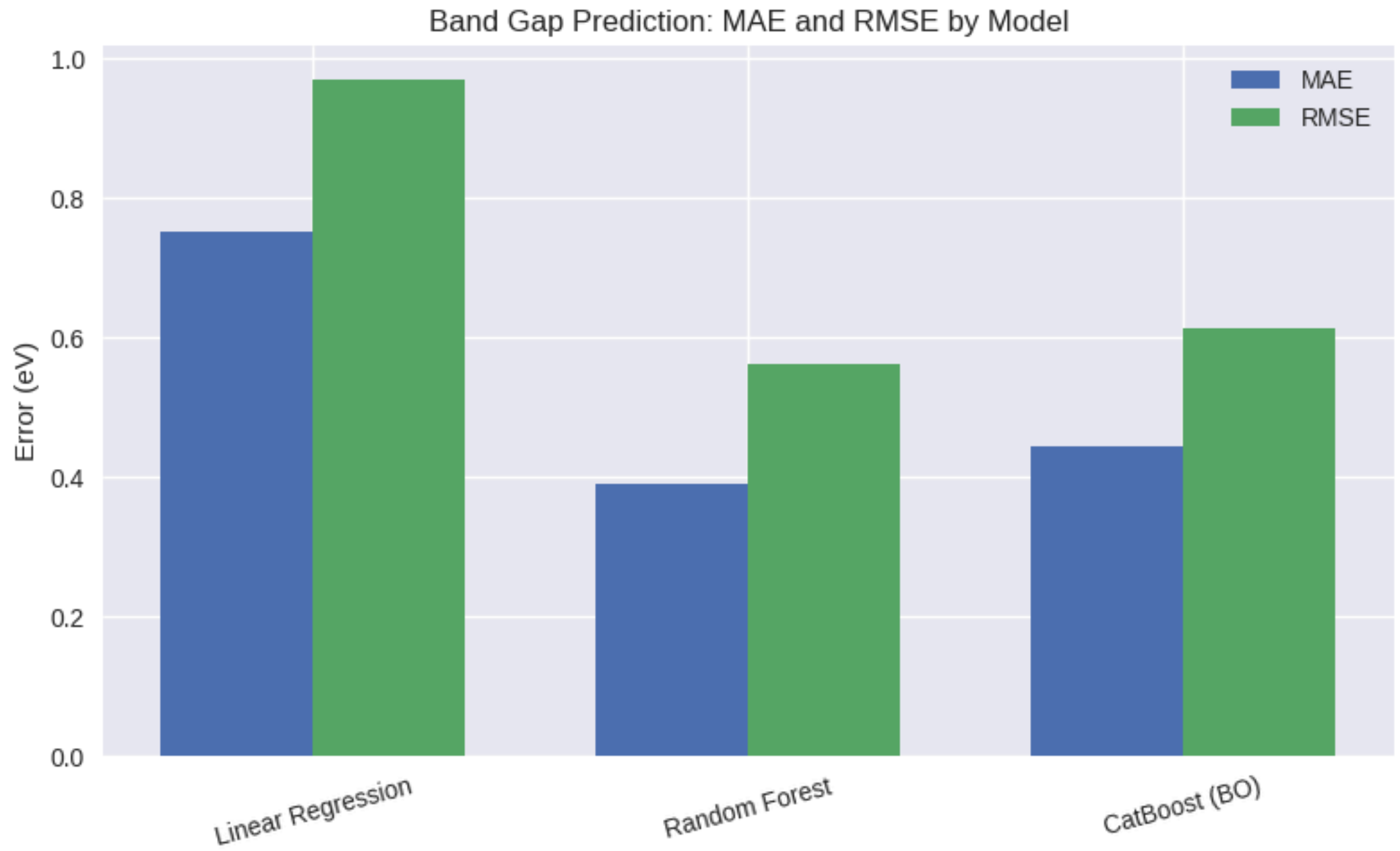

*Figure 4. Comparison of MAE and RMSE for Linear Regression, Random Forest, and Bayesian-optimised CatBoost models.*

Parity and residual plots for the Bayesian-optimised CatBoost model show that its predictions followed the 1:1 line over most of the range and show good agreement, with systematic under-prediction only at the largest band gaps, which were sparsely represented (Figure 5). Residuals were centered near 0 eV and show no clear pattern across most of the domain, except for an expected increase in error at the high band gap tail region, where data density was again low (Figure 6). For early-stage catalyst screening purposes, this level of performance is still valuable since the model reliably separates metallic and narrow band gap materials from moderate- to wide- band gap candidates and can therefore prioritize which structures warrant further experimental synthesis and analysis.

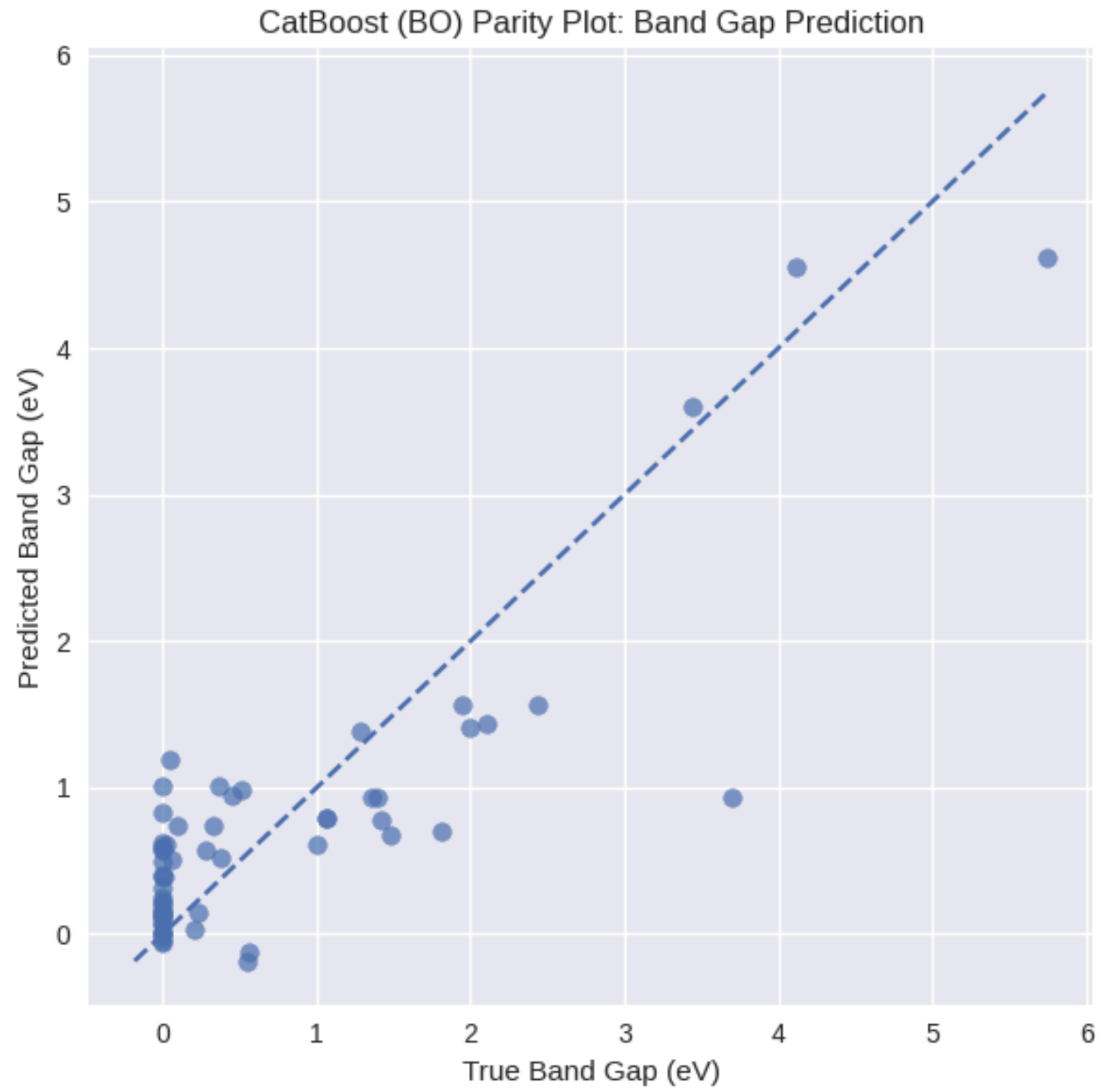

*Figure 5. Parity plot for the Bayesian-optimised CatBoost model, predicted versus true band gaps.*

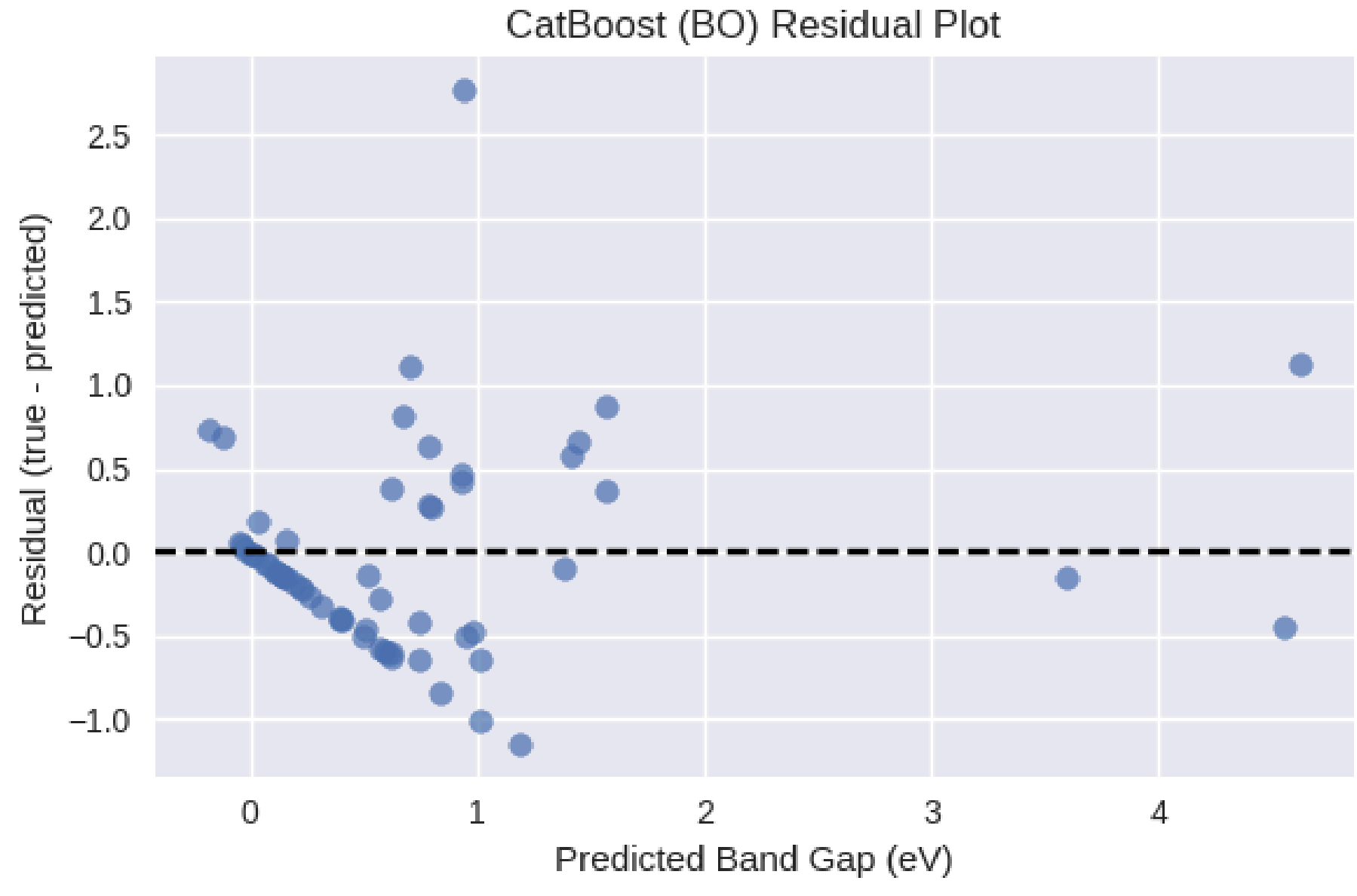

*Figure 6. Residual plot for the Bayesian-optimised CatBoost model, residuals (true - predicted band gap) versus predicted band gap.*

From a modeling standpoint, these trends follow expected bias-variance behavior: Linear regression underfits because it imposes a high-bias assumption that all relationships are linear combinations of the inputs, Random Forest lowers this bias by capturing non-linear interactions through its recursive splits from its decision tree ensemble, and CatBoost builds on this with boosted trees that capture more subtle dependencies which can be inherent in these diverse microdescriptors in the relatively small dataset. The aforementioned error and variance metrics confirm that the structure-property landscape in the catalytic microenvironment is non-linear and that non-linear tree-based methods are much more appropriate tools for navigating this complexity. In this particular dataset, Random Forest achieves the strongest numerical performance compared to Linear Regression and boosting algorithms and this can be expected for a small, low-dimensional, and relatively clean materials datasets. Importantly, the Bayesian-optimized CatBoost model remains valuable in this workflow because its predictions are smoother and its boosting structure provides clearer interpretability for SHAP analysis.

### *3.4 Feature Importance and SHAP Interpretability*

The Random Forest feature importance analysis reveals a clear hierarchy in how the microdescriptors influence band gap prediction: the formation energy per atom is overwhelmingly dominant, accounting for more than half of the total importance (~0.57), followed by the related energy per atom accounting for under 20% of the total importance (~0.19), and then the stability-related energy above hull accounting for under 6% of the total importance (~0.055) (Figure 7). More minor contributions come from the geometry- or size-related microdescriptors like the density, volume, and the number of sites while the compositional or stoichiometric fractions and ratios have very small influence. This ranking is chemically intuitive since more strongly bonded or stabilized material lattices typically show wider band gaps while less stabilized or distorted structures tend to show more metallic or near-metallic behavior. The geometric microdescriptors serve as proxies for material framework openness and the coordination environment which affects the d-orbital overlap and band. The much lower importance of the composition or stoichiometric fractions indicates that, within this Fe/Si/Al/O space, the electronic structure is more governed by how the atoms are arranged and stabilized in 3D space than just by their elemental counts.

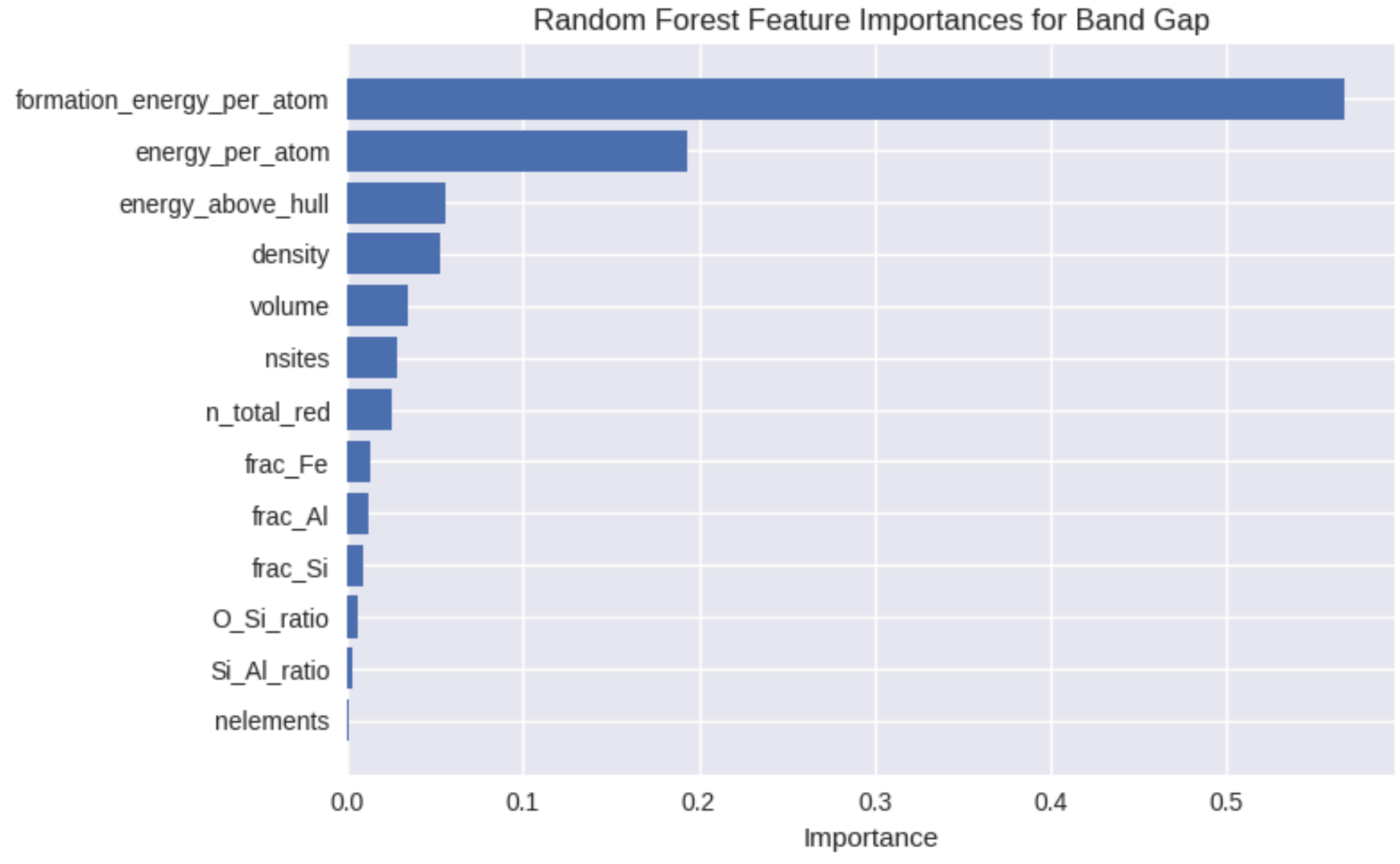

*Figure 7. Random Forest feature importance ranking for band gap prediction.*

SHAP analysis of the Bayesian-optimised CatBoost model largely confirms and refines these same trends, with the SHAP global feature importance ranking showing that band gap formation in the Fe/Si/Al/O space is governed primarily by thermodynamic stability microdescriptors like formation energies and energy above the hull collectively far outweighing all other features (Figure 8). With some nuance or variation, the next tier of microdescriptors again includes geometric or size-specific microdescriptors like volume, density, and the number of sites, elements, and atoms in the reduced formula of the structure. Stoichiometric and compositional fractions continue to contribute little to the prediction of electronic behavior. The feature importance hierarchies from both non-linear tree-based models thus indicate that electronic structure depends more on how the atoms are arranged and stabilized in 3D space than just by their elemental proportions in the Fe-oxide and -zeolite system.

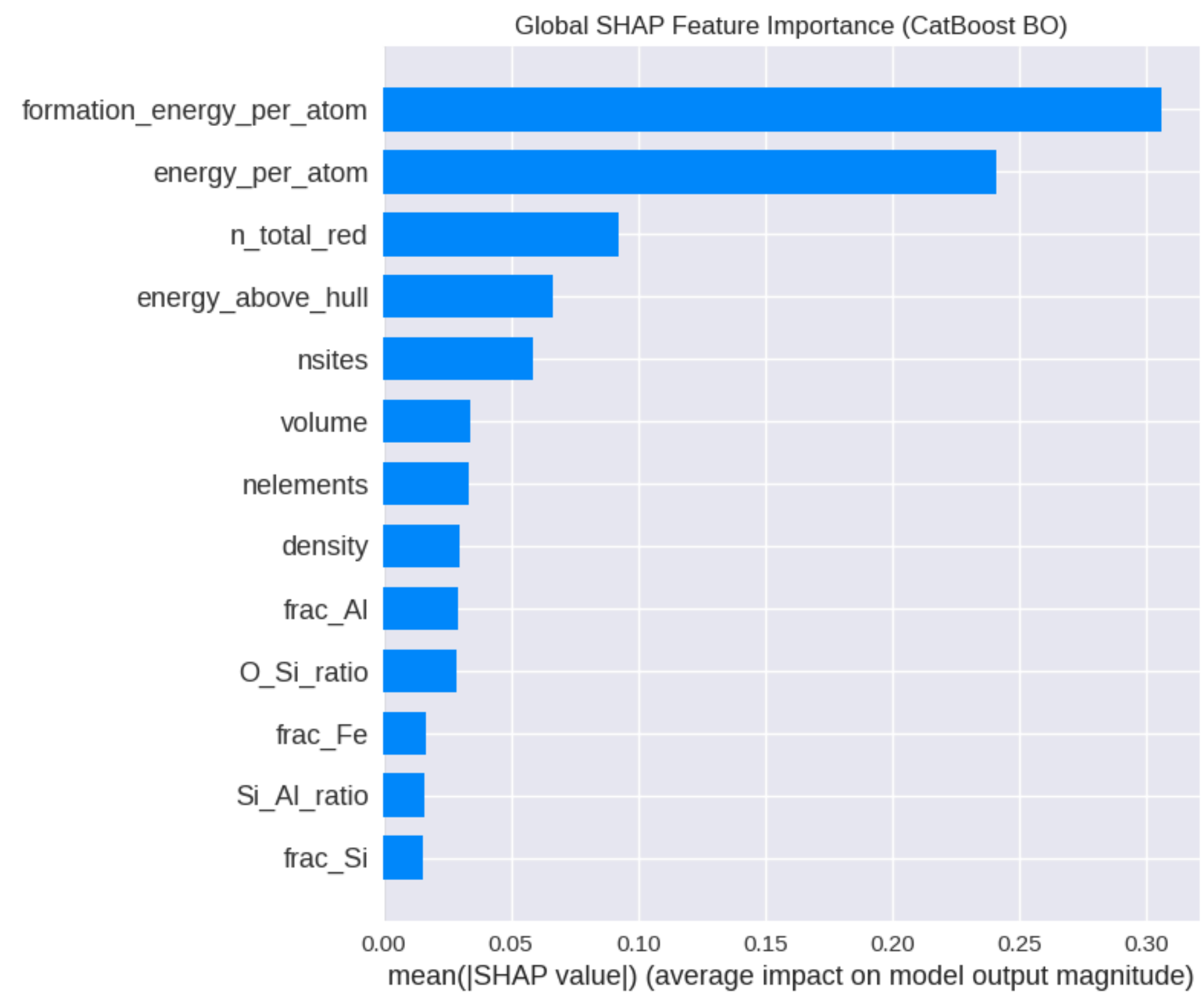

*Figure 8. Bayesian-optimized CatBoost feature importance ranking for band gap prediction.*

The SHAP global rankings identify which microdescriptors matter most across the dataset and the SHAP dependence plots show how each variable influences target variable predictions. Similar to the feature importance ranking, the global SHAP summary plot for the Bayesian-optimised CatBoost model shows that formation energy per atom has the largest spread of SHAP values, followed by energy per atom, and other microdescriptors (Figure 9). This means that variations in lattice stability and overall framework size drive the largest positive or negative contributions to the predicted band gap, a conclusion also derived from the feature importance rankings described above. The SHAP dependence plot for the highest-ranked feature/microdescriptor, the formation energy per atom, clearly shows a monotonically decreasing and essentially inverse relationship: more negative formation energies yield more positive SHAP values and therefore increase the predicted band gap, while less negative formation energies reduce the band gap (Figure 10). This pattern aligns with the mechanistic intuition also described above, where stronger bonding and higher lattice stability tend to widen the electronic band gap, while less stable or distorted frameworks support smaller gaps and more metallic behavior.

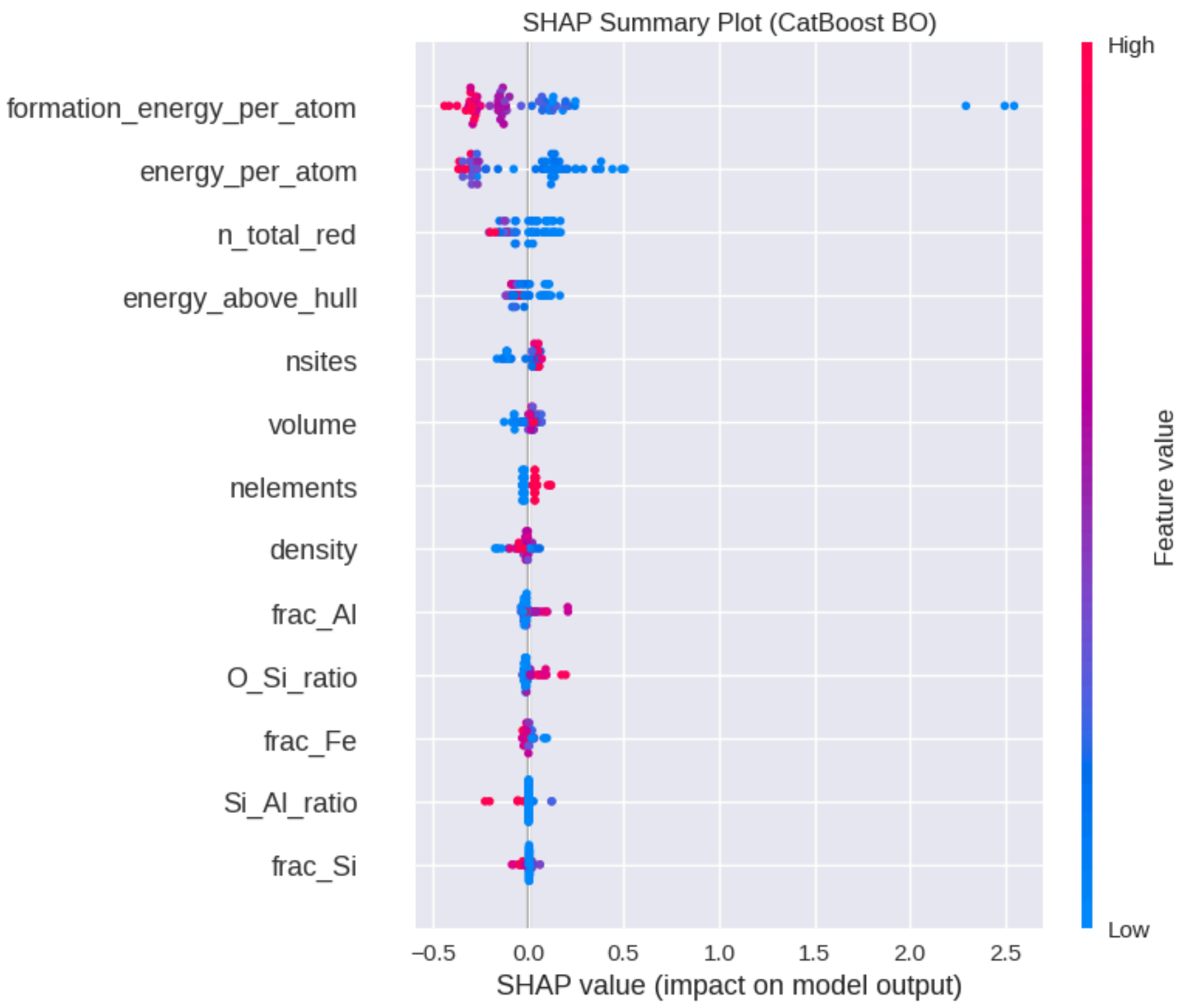

*Figure 9. SHAP summary plot for the Bayesian-optimised CatBoost model.*

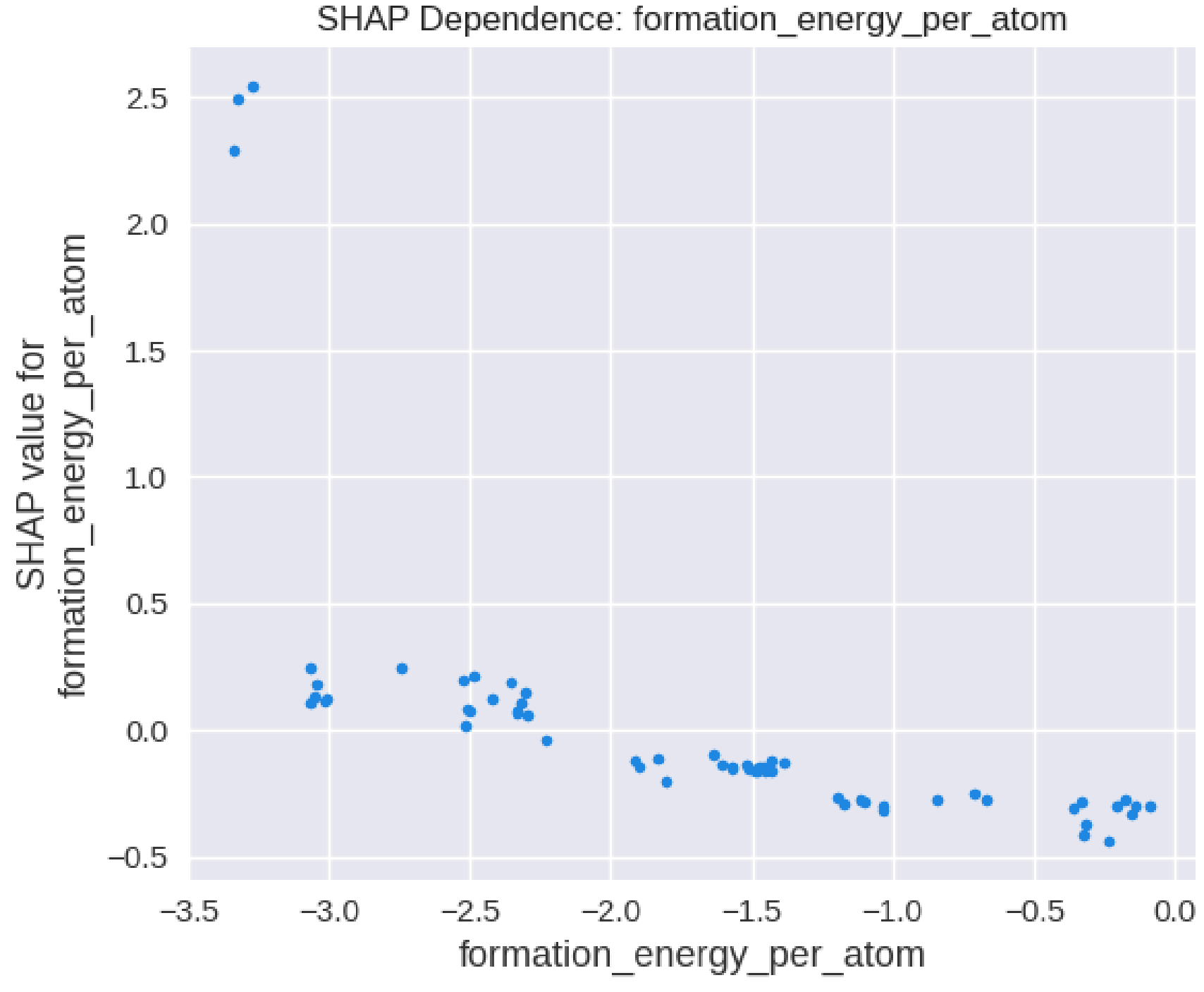

*Figure 10. SHAP dependence plot for formation energy per atom.*

Overall, SHAP provides a transparent, quantitative explanation for why stability-driven and geometry-driven microdescriptors like thermodynamic energies and lattice structure metrics dominate band gap formation in this Fe/Si/Al/O chemical family and that composition alone is insufficient to explain electronic behavior. These findings provide mechanistic understanding to a complex black-box system and an important link between supervised learning and catalyst design by showing that even crude microdescriptors can capture physically meaningful and chemically intuitive structure-electronic relationships.

### *3.5 Limitations and Future Extensions of Current Work for POM*

This study provided a clear first step toward more data-driven catalyst screening, but some key limitations must be acknowledged:

- The dataset was modest in size and restricted to Fe/Si/Al/O structures already present in the Materials Project database. As a result, the entire composition and topology space relevant to methane partial oxidation was only partially sampled and the current trained models may not necessarily extend to the defect-rich microenvironments encountered in real catalysts.
- Although the target variable, band gap, is an electronic property and a reasonable proxy for redox flexibility or reactivity and in some non-linear ways, the d-band center, it does not explicitly encode adsorption barriers, reaction energetics, or diffusion limitations. Thus, the present models are electronic structure-microdescriptor surrogates rather than full predictors of POM catalytic conversion, selectivity, or yield.

Despite these limitations, band gap remains a meaningful microdescriptor for Fe-based POM catalysis because it governs Fe oxidation-state flexibility, electron transfer to $O_2$ and $CH_4$, and the stability of radical oxygen species, which are processes that directly influence conversion and selectivity. The fact that a small set of thermodynamic and structural microdescriptors explains most of the variance in band gap indicates that similar surrogate models could be developed for adsorption energies, activation barriers, and other properties more directly tied to kinetic performance. Thus, a natural next step for future work, asides from further hyperparameter tuning of the non-linear tree-based towards improved optimization and potentially incorporating other state-of-the-art ML techniques for robustness, is to link the present electronic structure-microdescriptor surrogate models to microkinetic models and subsequently reaction engineering models. Therein, the machine-learned band gaps (or other electronic microdescriptors) could be mapped to approximate activation energies using scaling relations or small DFT calibration sets and these activation energies would then dictate pre-exponential (e.g., Arrhenius) rate constants in a microkinetic scheme and enable the computation of reactant conversion and product selectivity through reaction engineering equations.

More broadly, this robust pipeline would transform catalytic structure-property relationships into reactor-level, commercially-relevant performance metrics that support future integration into digital twin reactor engineering models that can intelligently narrow vast design spaces and reduce the cost of hypothesis testing towards interpretable insights. Effectively, this pipeline would allow catalytic insights to propagate through the entire reactor design and processing engineering workflow since reactor mass & energy balances, output temperatures and heat duties, product composition profiles, and even recycle ratios and energy efficiency metrics all tie back to reactant conversion and selectivity, which are fundamentally catalytic-dependent parameters. Furthermore, the combination of such digital twin models with layers of technoeconomic analysis (i.e., TEA) or financial modeling in turn allow for these reactor-level outputs to determine key TEA or financial/business metrics such as cost per tonne of product or product yield, carbon intensity, OPEX/CAPEX scaling, and general economic feasibility.

## 4. Conclusion

Catalyst discovery for commercially-relevant engineered systems, such as for methane partial oxidation (i.e., POM), remains fundamentally constrained by slow, costly, and intuition-dependent experimentation. The size of the catalyst design space and the complexity inherent in the interplay of microenvironmental catalyst features like electronic structure and adsorption energies makes this traditional trial-and-error R&D approach inefficient and difficult to scale not just within a singular catalysis development study, but also more generally to broader catalysis systems. This current work demonstrated that supervised learning models trained on the curated yet crude structural, thermodynamic, and compositional microdescriptor information from open catalyst property datasets can reveal meaningful structure-performance relationships that would be too arduous to extract traditionally. An interpretable, non-linear tree-based machine learning framework that predicts band gaps in Fe/Si/Al/O catalytic materials for the POM reaction system was developed: Random Forest and Bayesian-optimized CatBoost models achieved strong predictive performance against linear regression models and SHAP interpretation or explainability analysis reveals that band gap is governed primarily by the thermodynamic lattice stabilization and geometric factors rather than bulk stoichiometry or composition of the catalyst material. The aforementioned insights are chemically consistent and align with known mechanistic and materials science principles and reveal a ranked feature importance list of which catalyst attributes deserve more early-stage experimental focus. The methodology demonstrated within this work can be extended to adsorption energy and activation barrier analysis, thus establishing a transferable and extendable method for interpretability and the scalable evaluation of catalytic microenvironments. In all, this work forms the early architectural foundation of a closed-loop, AI-guided R&D system and, in the medium- to long-term, can allow for the development and proliferation of physically- and chemically-grounded "world models" that can be used in or as precursors to unified, self-driving R&D laboratories that are capable of actively and autonomously proposing/recommending, executing, and interpreting experiments, updating the catalyst-property models with fresh data, and recalculating process economics in real time. Effectively, it is an initial step in driving the next-generation of energy, chemical, and materials innovation in engineering and applied science applications of urgent importance.

## 5. References (IEEE)

[1] Z. Ma and F. Zaera, “Heterogeneous Catalysis by Metals,” in *Encyclopedia of Inorganic and Bioinorganic Chemistry*, John Wiley & Sons, Ltd, 2014, pp. 1–16. doi: 10.1002/9781119951438.eibc0079.pub2.

[2] “Introduction to Surface Chemistry and Catalysis, 2nd Edition | Wiley,” Wiley.com. [Online]. Available: https://www.wiley.com/en-us/Introduction+to+Surface+Chemistry+and+Catalysis%2C+2nd+Edition-p-9780470508237

[3] A. Abbas *et al.*, “Catalysis at the intersection of sustainable chemistry and a circular economy,” *One Earth*, vol. 7, no. 5, pp. 738–741, May 2024, doi: 10.1016/j.oneear.2024.04.018.

[4] A. A. Latimer, A. Kakekhani, A. R. Kulkarni, and J. K. Nørskov, “Direct Methane to Methanol: The Selectivity–Conversion Limit and Design Strategies,” *ACS Catal.*, vol. 8, no. 8, pp. 6894–6907, Aug. 2018, doi: 10.1021/acscatal.8b00220.

[5] J. K. Nørskov *et al.*, “Universality in Heterogeneous Catalysis,” *J. Catal.*, vol. 209, no. 2, pp. 275–278, July 2002, doi: 10.1006/jcat.2002.3615.

[6] “The Genesis of Molecular Volcano Plots | Accounts of Chemical Research.” [Online]. Available: https://pubs.acs.org/doi/10.1021/acs.accounts.0c00857

[7] B. Hammer and J. K. Nørskov, “Electronic factors determining the reactivity of metal surfaces,” *Surf. Sci.*, vol. 343, no. 3, pp. 211–220, Dec. 1995, doi: 10.1016/0039-6028(96)80007-0.

[8] B. Wang *et al.*, “Strain engineering of high-entropy alloy catalysts for electrocatalytic water splitting,” *iScience*, vol. 26, no. 4, p. 106326, Apr. 2023, doi: 10.1016/j.isci.2023.106326.

[9] L. C. Grabow and M. Mavrikakis, “Mechanism of Methanol Synthesis on Cu through CO2 and CO Hydrogenation,” *ACS Catal.*, vol. 1, no. 4, pp. 365–384, Apr. 2011, doi: 10.1021/cs200055d.

[10] Z.-K. Han, W. Liu, and Y. Gao, “Advancing the Understanding of Oxygen Vacancies in Ceria: Insights into Their Formation, Behavior, and Catalytic Roles,” *JACS Au*, vol. 5, no. 4, pp. 1549–1569, Apr. 2025, doi: 10.1021/jacsau.5c00095.

[11] J. H. Lunsford, “Catalytic conversion of methane to more useful chemicals and fuels: a challenge for the 21st century,” *Catal. Today*, vol. 63, no. 2, pp. 165–174, Dec. 2000, doi: 10.1016/S0920-5861(00)00456-9.

[12] N. Naeem *et al.*, “Recent developments and discourse on catalyst systems for partial oxidation of methane to syngas production,” *Int. J. Hydrog. Energy*, vol. 139, pp. 646–717, June 2025, doi: 10.1016/j.ijhydene.2025.05.183.

[13] S. Kim, S. Nam, W. Jung, H. Kim, Y. Choi, and H. Kim, “Designing Highly Active and Stable Ni-Exsolved LaMnO3 Perovskite Catalysts for Dry Reforming of Methane via Ca Substitution,” *ACS Catal.*, vol. 15, no. 9, pp. 6812–6825, May 2025, doi: 10.1021/acscatal.5c00570.

[14] B. Christian Enger, R. Lødeng, and A. Holmen, “A review of catalytic partial oxidation of methane to synthesis gas with emphasis on reaction mechanisms over transition metal catalysts,” *Appl. Catal. Gen.*, vol. 346, no. 1, pp. 1–27, Aug. 2008, doi: 10.1016/j.apcata.2008.05.018.

[15] R. Horn and R. Schlögl, “Methane Activation by Heterogeneous Catalysis,” *Catal. Lett.*, vol. 145, no. 1, pp. 23–39, Jan. 2015, doi: 10.1007/s10562-014-1417-z.

[16] P. P. Knops-Gerrits and W. A. Goddard, “Methane partial oxidation in iron zeolites: theory versus experiment,” *J. Mol. Catal. Chem.*, vol. 166, no. 1, pp. 135–145, Jan. 2001, doi: 10.1016/S1381-1169(00)00460-X.

[17] J. A. Esterhuizen, B. R. Goldsmith, and S. Linic, “Theory-Guided Machine Learning Finds Geometric Structure-Property Relationships for Chemisorption on Subsurface Alloys,” *Chem*, vol. 6, no. 11, pp. 3100–3117, Nov. 2020, doi: 10.1016/j.chempr.2020.09.001.

[18] M. J. Nielsen, L. H. E. Kempen, J. de Neergaard Ravn, R. Cheula, and M. Andersen, “Interpretable machine learned predictions of adsorption energies at the metal–oxide interface,” *J. Chem. Phys.*, vol. 163, no. 4, p. 044708, July 2025, doi: 10.1063/5.0282674.

[19] L. Zuo, P. Ni, T. Tanaka, and Y. Li, "Machine Learning on Contact Angles of Liquid Metals and Solid Oxides," *Metall. Mater. Trans. B*, vol. 52, no. 1, pp. 17–22, Feb. 2021, doi: 10.1007/s11663-020-02013-5.
[20] A. Jain *et al.*, "Commentary: The Materials Project: A materials genome approach to accelerating materials innovation," *APL Mater.*, vol. 1, no. 1, p. 011002, July 2013, doi: 10.1063/1.4812323.
[21] L. Chanussot *et al.*, "The Open Catalyst 2020 (OC20) Dataset and Community Challenges," *ACS Catal.*, vol. 11, no. 10, pp. 6059–6072, May 2021, doi: 10.1021/acscatal.0c04525.
[22] "Catalysis-Hub.org, an open electronic structure database for surface reactions | Scientific Data." [Online]. Available: https://www.nature.com/articles/s41597-019-0081-y
[23] "NIST Chemical Kinetics Database." [Online]. Available: https://kinetics.nist.gov/kinetics/
[24] Z. W. Ulissi, A. J. Medford, T. Bligaard, and J. K. Nørskov, "To address surface reaction network complexity using scaling relations machine learning and DFT calculations," *Nat. Commun.*, vol. 8, p. 14621, Mar. 2017, doi: 10.1038/ncomms14621.
[25] M. Suvarna, T. Zou, S. H. Chong, Y. Ge, A. J. Martín, and J. Pérez-Ramírez, "Active learning streamlines development of high performance catalysts for higher alcohol synthesis," *Nat. Commun.*, vol. 15, no. 1, p. 5844, July 2024, doi: 10.1038/s41467-024-50215-1.
[26] R. Gómez-Bombarelli *et al.*, "Automatic Chemical Design Using a Data-Driven Continuous Representation of Molecules," *ACS Cent. Sci.*, vol. 4, no. 2, pp. 268–276, Feb. 2018, doi: 10.1021/acscentsci.7b00572.
[27] Z. Rao *et al.*, "Machine learning–enabled high-entropy alloy discovery," *Science*, vol. 378, no. 6615, pp. 78–85, Oct. 2022, doi: 10.1126/science.abo4940.
[28] L. Mencarelli, A. Pagot, and P. Duchêne, "Surrogate-based modeling techniques with application to catalytic reforming and isomerization processes," *Comput. Chem. Eng.*, vol. 135, p. 106772, Apr. 2020, doi: 10.1016/j.compchemeng.2020.106772.
[29] O. Romiluyi, Y. Eatmon, R. Ni, B. P. Rand, and P. Clancy, "The efficacy of Lewis affinity scale metrics to represent solvent interactions with reagent salts in all-inorganic metal halide perovskite solutions," *J. Mater. Chem. A*, vol. 9, no. 22, pp. 13087–13099, June 2021, doi: 10.1039/D1TA03063A.
[30] B. P. MacLeod *et al.*, "Self-driving laboratory for accelerated discovery of thin-film materials," *Sci. Adv.*, vol. 6, no. 20, p. eaaz8867, May 2020, doi: 10.1126/sciadv.aaz8867.
[31] N. J. Szymanski *et al.*, "An autonomous laboratory for the accelerated synthesis of novel materials," *Nature*, vol. 624, no. 7990, pp. 86–91, Dec. 2023, doi: 10.1038/s41586-023-06734-w.
[32] F. Häse, L. M. Roch, and A. Aspuru-Guzik, "Next-Generation Experimentation with Self-Driving Laboratories," *Trends Chem.*, vol. 1, no. 3, pp. 282–291, June 2019, doi: 10.1016/j.trechm.2019.02.007.
[33] R. Astudillo, D. R. Jiang, M. Balandat, E. Bakshy, and P. I. Frazier, "Multi-Step Budgeted Bayesian Optimization with Unknown Evaluation Costs," Nov. 12, 2021, *arXiv*: arXiv:2111.06537. doi: 10.48550/arXiv.2111.06537.
[34] P. I. Frazier, "A Tutorial on Bayesian Optimization," July 08, 2018, *arXiv*: arXiv:1807.02811. doi: 10.48550/arXiv.1807.02811.
[35] A. Talapatra *et al.*, "Experiment Design Frameworks for Accelerated Discovery of Targeted Materials Across Scales," *Front. Mater.*, vol. 6, Apr. 2019, doi: 10.3389/fmats.2019.00082.
[36] T. Lookman, P. V. Balachandran, D. Xue, and R. Yuan, "Active learning in materials science with emphasis on adaptive sampling using uncertainties for targeted design," *Npj Comput. Mater.*, vol. 5, no. 1, p. 21, Feb. 2019, doi: 10.1038/s41524-019-0153-8.
[37] J. K. Pedersen *et al.*, "Bayesian Optimization of High-Entropy Alloy Compositions for Electrocatalytic Oxygen Reduction," *Angew. Chem. Int. Ed.*, vol. 60, no. 45, pp. 24144–24152, 2021, doi: 10.1002/anie.202108116.
[38] Y. Kwon, D. Lee, J. W. Kim, Y.-S. Choi, and S. Kim, "Exploring Optimal Reaction Conditions Guided by Graph Neural Networks and Bayesian Optimization," *ACS Omega*, vol. 7, no. 49, pp. 44939–44950, Dec. 2022, doi: 10.1021/acsomega.2c05165.

[39] B. J. Shields *et al.*, “Bayesian reaction optimization as a tool for chemical synthesis,” *Nature*, vol. 590, no. 7844, pp. 89–96, Feb. 2021, doi: 10.1038/s41586-021-03213-y.

[40] F. Häse, L. M. Roch, C. Kreisbeck, and A. Aspuru-Guzik, “Phoenics: A Bayesian Optimizer for Chemistry,” *ACS Cent. Sci.*, vol. 4, no. 9, pp. 1134–1145, Sept. 2018, doi: 10.1021/acscentsci.8b00307.

[41] S. Lundberg and S.-I. Lee, “A Unified Approach to Interpreting Model Predictions,” Nov. 25, 2017, *arXiv*: arXiv:1705.07874. doi: 10.48550/arXiv.1705.07874.

[42] G. Pilania, “Machine learning in materials science: From explainable predictions to autonomous design,” *Comput. Mater. Sci.*, vol. 193, p. 110360, June 2021, doi: 10.1016/j.commatsci.2021.110360.

[43] X. Li *et al.*, “Sequential closed-loop Bayesian optimization as a guide for organic molecular metallophotocatalyst formulation discovery,” *Nat. Chem.*, vol. 16, no. 8, pp. 1286–1294, Aug. 2024, doi: 10.1038/s41557-024-01546-5.

[44] A. Nandy, C. Duan, M. G. Taylor, F. Liu, A. H. Steeves, and H. J. Kulik, “Computational Discovery of Transition-metal Complexes: From High-throughput Screening to Machine Learning,” *Chem. Rev.*, vol. 121, no. 16, pp. 9927–10000, Aug. 2021, doi: 10.1021/acs.chemrev.1c00347.

[45] M. T. Gorzkowski and A. Lewera, “Probing the Limits of d-Band Center Theory: Electronic and Electrocatalytic Properties of Pd-Shell–Pt-Core Nanoparticles,” *J. Phys. Chem. C*, vol. 119, no. 32, pp. 18389–18395, Aug. 2015, doi: 10.1021/acs.jpcc.5b05302.

[46] S. Tang, W. Shuang, Y. Wu, Z. Jia, Z. Bai, and L. Yang, “3d-orbital overlap modulated d-band center of high-entropy oxyhydroxide for efficient oxygen evolution reaction,” *Appl. Surf. Sci.*, vol. 682, p. 161760, Feb. 2025, doi: 10.1016/j.apsusc.2024.161760.

[47] J. Zhao, W. Xia, Z. Zeng, and X. Wang, “Central role of d-band energy level in Cu-based intermetallic alloys,” *Npj Comput. Mater.*, vol. 10, no. 1, p. 71, Apr. 2024, doi: 10.1038/s41524-024-01257-y.

[48] R. Murase, C. F. Leong, and D. M. D’Alessandro, “Mixed Valency as a Strategy for Achieving Charge Delocalization in Semiconducting and Conducting Framework Materials,” *Inorg. Chem.*, vol. 56, no. 23, pp. 14373–14382, Dec. 2017, doi: 10.1021/acs.inorgchem.7b02090.

[49] M. A. Syzgantseva, C. P. Ireland, F. M. Ebrahim, B. Smit, and O. A. Syzgantseva, “Metal Substitution as the Method of Modifying Electronic Structure of Metal–Organic Frameworks,” *J. Am. Chem. Soc.*, vol. 141, no. 15, pp. 6271–6278, Apr. 2019, doi: 10.1021/jacs.8b13667.

[50] J. A. Branson *et al.*, “The Counterintuitive Relationship between Orbital Energy, Orbital Overlap, and Bond Covalency in CeF62– and CeCl62–,” *J. Am. Chem. Soc.*, vol. 146, no. 37, pp. 25640–25655, Sept. 2024, doi: 10.1021/jacs.4c07459.

[51] K. T. Butler, C. H. Hendon, and A. Walsh, “Electronic Structure Modulation of Metal–Organic Frameworks for Hybrid Devices,” *ACS Appl. Mater. Interfaces*, vol. 6, no. 24, pp. 22044–22050, Dec. 2014, doi: 10.1021/am507016r.